%% file: ctrl_paper.tex
\documentclass[journal]{IEEEtran}
\usepackage{amsmath,amsfonts}
\usepackage{algorithmic}
\usepackage{fontawesome5}

\usepackage{array}
\usepackage[caption=false,font=normalsize,labelfont=sf,textfont=sf]{subfig}
\usepackage{textcomp}
\usepackage{stfloats}
\usepackage{hhline}
\usepackage{url}
\usepackage{verbatim}
\usepackage{graphicx}
\def\BibTeX{{\rm B\kern-.05em{\sc i\kern-.025em b}\kern-.08em
    T\kern-.1667em\lower.7ex\hbox{E}\kern-.125emX}}
\usepackage{balance}

\usepackage[table,dvipsnames]{xcolor}
\usepackage[shrink=25]{microtype}
\usepackage[mathscr]{eucal}
\usepackage{dsfont}
\usepackage{scalefnt}
\usepackage{amsmath,amssymb,amsfonts}
\usepackage{algorithmic}
\usepackage{graphicx}
\usepackage{textcomp}
\usepackage{listings}
\usepackage{multirow}
\usepackage{pgfplotstable}
\usepackage{pgfplots}
\usepackage{marvosym}
\usepackage[english]{babel}
\usepackage{upgreek}
\usepackage{makecell}
\usepackage{framed}
\usepackage{fp}
\usepackage{setspace}
\usepackage{pifont}
\usepackage[nolist]{acronym}
\def\BibTeX{{\rm B\kern-.05em{\sc i\kern-.025em b}\kern-.08em
    T\kern-.1667em\lower.7ex\hbox{E}\kern-.125emX}}

\usepackage[hidelinks]{hyperref}
\usepackage[backend=biber, style=numeric, hyperref, maxnames=1, minnames=1, giveninits=true, maxcitenames=2, mincitenames=1,bibencoding=utf8, isbn=true,sorting=none]{biblatex}
\usepackage{cleveref}
\usepackage{siunitx}
\usepackage{xspace}
\usepackage{xargs}

\definecolor{codegreen}{rgb}{0,0.6,0}
\definecolor{codegray}{rgb}{0.5,0.5,0.5}
\definecolor{codepurple}{rgb}{0.58,0,0.82}
\definecolor{backcolour}{rgb}{0.95,0.95,0.92}

\lstdefinestyle{mystyle}{
    backgroundcolor=\color{backcolour},
    commentstyle=\color{codegreen},
    keywordstyle=\bfseries\color{magenta},
    numberstyle=\tiny\color{codegray},
    stringstyle=\color{codepurple},
    basicstyle=\ttfamily\footnotesize,
    breakatwhitespace=false,
    breaklines=true,
    captionpos=b,
    keepspaces=true,
    numbers=left,
    numbersep=5pt,
    showspaces=false,
    showstringspaces=false,
    showtabs=false,
    tabsize=1
}

\usepackage{csquotes}
\newcommand{\headingBlock}[1]{\par\noindent{#1}~}

\DeclareSIUnit\flops{FLOPS}
\DeclareSIUnit\flop{FLOP}

\DeclareSIUnit\ops{OPS}
\DeclareSIUnit\op{OP}

\pgfplotsset{compat=1.18}

\graphicspath{{figures/}}

\input{include.tex}

\DeclareSIUnit\flops{FLOPS}
\DeclareSIUnit\flop{FLOP}
\DeclareSIUnit\luts{LUTs}
\DeclareSIUnit\ffs{FFs}
\DeclareSIUnit\brams{BRAMs}
\DeclareSIUnit\rambhs{RAMB18s}
\DeclareSIUnit\rambfs{RAMB36s}
\DeclareSIUnit\rambh{RAMB18}
\DeclareSIUnit\rambf{RAMB36}

\DeclareSIUnit\ops{OPS}
\DeclareSIUnit\op{OP}

\begin{document}

\newcommand{\eg}{e.\,g.}
\newcommand{\ie}{i.\,e.}
\definecolor{cmark_green}{HTML}{00A64F}
\newcommand{\cmark}{\textcolor{cmark_green}{\ding{51}}}
\definecolor{xmark_red}{HTML}{ED1B23}
\newcommand{\xmark}{\textcolor{xmark_red}{\ding{55}}}
\definecolor{mmark_yellow}{HTML}{ED7A1B}
\newcommand{\mmark}{\textcolor{mmark_yellow}{\ding{51}}}

\newcolumntype{C}[1]{>{\centering\let\newline\\\arraybackslash\hspace{0pt}}m{#1}}

\newcommand{\secRef}[1]{Section~\ref{#1}}
\newcommand{\algoRef}[1]{Algorithm~\ref{#1}}
\newcommand{\codeRef}[1]{Listing~\ref{#1}}
\newcommand{\figRef}[1]{Figure~\ref{#1}}
\newcommand{\tableRef}[1]{Table~\ref{#1}}
\setlength\arraycolsep{1pt}
\newcommand{\praEqRef}[1]{(Eq.~$S_{#1}$)}

\newcommand{\factor}[1]{\qty{#1}{\times}}
\newcommand{\factorRange}[2]{\qtyrange{#1}{#2}{\times}}

\newcommand{\memSize}[2]{#1$\times$\qty{#2}{\bit}}


\title{Loop Control Management in\\ Tightly Coupled Processor Arrays~(TCPAs)}

\author{
	\IEEEauthorblockN{Dominik Walter, Frank Hannig, J\"urgen Teich\\}
	\IEEEauthorblockA{cs12-alpaca@fau.de\\
	Hardware/Software Co-Design, Department of Computer Science\\
	Friedrich-Alexander-Universit\"at Erlangen-N\"urnberg (FAU), Germany
			}
}

\maketitle

\begin{abstract}
Multidimensional loop kernels often suffer from control overhead that can dominate execution time on parallel loop accelerators.
Tightly Coupled Processor Arrays (TCPAs) offload loop control to a global controller (GC), but existing approaches still require hundreds of control signals.
We propose a method to derive and aggressively reduce these control conditions from a polyhedral representation of the iteration space, achieving reductions of $\factorRange{15}{45}$ in control signals across several benchmarks.
We introduce a lightweight GC architecture that evaluates conditions as unions of polyhedra using bounded evaluation units, requiring hardware comparable to a single processing element.
Control signals are distributed throughout the array with a minimal number of delay elements resulting in zero-overhead loop control.
Our evaluation on PolyBench kernels shows that the entire control flow requires $<\qty{10}{\percent}$ of the total array resources.
\end{abstract}


\section{Introduction}
    It is well known that, in many applications, the overhead required for control can amount to \qty{75}{\percent} or more of total execution time.
    This is especially the case for multidimensional nested loops common in linear algebra, machine learning, and AI.
    In recent years, the computational demands of these domains have grown rapidly, requiring sophisticated accelerator architectures, \eg, \cite{WangKMMP19, alpaca}, to keep pace.
    Therefore, a crucial aspect of designing such accelerators is reducing the control overhead as much as possible.
    A common class of loop accelerators is processor arrays, \ie, an array of small processor elements exploiting the inherent parallelism of the loop.
    An example of such an architecture is a \textit{Tightly Coupled Processor Array}~(TCPA).
    However, a TCPA can easily be crippled if the control overhead is not carefully considered in both software and hardware.
    Note that there are typically two types of control overhead in loops: loop control~(\eg, updating loop indices and conditions) and IO-related control~(\eg, loads, stores, address computations).
    In this work, we propose a solution to the first type of overhead, loop control, for TCPAs.
    More specifically, we
    \begin{itemize}
        \item Introduce TCPAs in more detail in \secRef{sec:tcpa},
        \item Compare this work to related architectures in \secRef{sec:related},
        \item Introduce the assignment, generation, propagation, and use of control signals that implement loop control in \secRef{sec:control}, and
        \item Evaluate the proposed approach in \secRef{sec:eval} on a set of loop benchmarks.
    \end{itemize}

        \begin{figure}
            \centering
            \resizebox{0.29\textwidth}{!}{
            \includegraphics[width=0.5\textwidth] {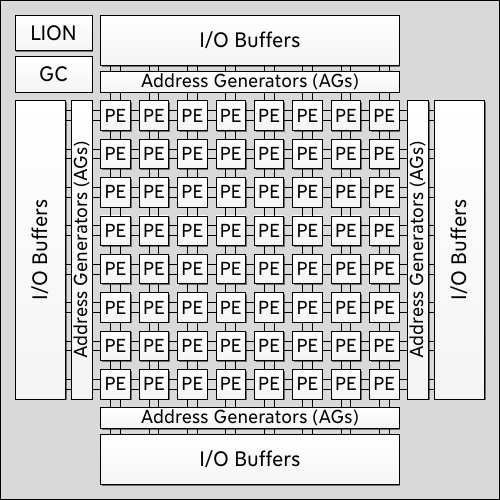}
            }
            \caption{Architecture of an $8 \times 8$ TCPA. The array of PEs is surrounded by 4 I/O buffers accessed by dedicated Address Generators~(AGs) and a Loop I/O Controller~(LION). All loop control is handled by the Global Controller~(GC).}
            \label{fig:tcpa_arch}
        \end{figure}
\section{Tightly Coupled Processor Arrays}
\label{sec:tcpa}
    This section briefly introduces the TCPA architecture and loop mapping.

    \subsection{Architecture}
        A TCPA~\cite{KisslerTeich2006IEEEFPT, HannigReiche2014ACMTECS, TeichWitterauf2022Book} is a two-dimensional array of Processing Elements~(PE), shown in \figRef{fig:tcpa_arch}.
        Each PE contains multiple Function Units (FUs) operating in parallel using Orthogonal Instruction Processing (OIP)~\cite{BrandTeich2017IEEEMCSOC}, \ie, each FU has its own tiny instruction memory containing an FU-specific microprogram.
        Each PE also contains a local register file that is shared by all FUs within the PE.
        The PEs are tightly coupled to their neighboring PEs via a switch-based interconnect network.
        At the border of the array, four I/O buffers provide input data.
        The I/O buffers are not accessed directly by PEs but by specialized Address Generators~(AGs).
        A Loop I/O Controller~(LION) dynamically fills and clears the I/O buffers at runtime by scheduling all required data transfers to/from an external memory~\cite{WalterT21}.
        Finally, \figRef{fig:tcpa_arch} also shows the Global Controller~(GC), which implements the central loop control of all PEs by generating control signals.

    \subsection{Mapping}
        TCPAs are designed to execute loops described as Piecewise Regular Algorithms~(PRAs)~\cite{Tei93}, where loop iterations are expressed by an $\plaDimension$-dimensional iteration space $\plaIterationSpaceDefinition$ with each element $\plaIterationLongDefinition$ denoting a single iteration.
        Operations within each iteration are defined by a set of equations $\plaEquationSpaceDefinition$ with:
        $$\plaEquationDefinition$$
        It defines for all iterations $\plaIteration \in \plaEquationDomain$ the variable $x_i$ at index $\plaWriteIndexingFunction$ as the result of applying operation $\plaOperation$ on variables $y_{i,j}$, where the condition space $\plaEquationDomain$ is given as a polyhedron:
        $$\plaConditionSpaceLongDefinition$$
        Such loops can be systematically mapped to TCPAs~\cite{Tei93, nelis1988, WitteraufTeich2016ASAP, TanaseHannig2018ATECS, WitteraufWHT21}.
        First, the iteration space $\plaIterationSpace$ is partitioned into $\plaTiledIterationSpace$: 
        $$\plaTiledIterationSpaceDefinition$$
        Each index $\plaInterIteration$ denotes a tile of intra-tile iterations $\plaIntraIteration$ mapped to a single PE.
        Thus, while intra-tile iterations run sequentially on a PE, all PEs operate in parallel.
        Next, all operations~$\plaOperation$ within an iteration must be scheduled by assigning each operation with latency $\plaScheduleOperationExecutionTime$ a start time $\plaScheduleOperationStartTime$ on a free FU such that all data dependences are satisfied.
        This defines the local latency~$\plaScheduleLocalLatency$ as:
        $$\plaScheduleLocalLatencyDefinition$$
        This schedule is repeated for each iteration, with start times determined by a linear schedule vector $\plaTiledScheduleVectorDefinition$ such that the start time $\plaScheduleOperationScheduleTime$ of operation~$\plaOperation$ in iteration $\plaTiledIteration$ is $\plaScheduleOperationScheduleTimeDefinition$.
        The initiation interval $\plaScheduleInitiationInterval$ denotes the number of cycles between consecutive iteration starts.
        Data dependences between operations are carried out by the local register file and interconnect.
        Each combination of operation $\plaOperation$ and required registers is encoded as an instruction, and the schedule defines for all FUs a sequence of instructions forming the microprograms stored in each FU's instruction memory.
        Since the condition spaces $\plaConditionSpace$ cause different instruction sequences for different iterations, these microprograms contain frequent branches depending statically on the current loop indices.

\section{Related Work}
\label{sec:related}
    Loop control is a fundamental challenge extending beyond TCPAs, affecting virtually all compute architectures.
    Traditional CPUs address this overhead using wide SIMD operations that process multiple data elements per instruction.
    GPUs take this further with SIMT, where thousands of lightweight cores execute the same instruction stream in parallel~\cite{lindholm2008nvidia, nickolls2008scalable}.
    The approach in TCPAs is fundamentally different from both CPU and GPU architectures.
    We do not rely on SIMD-style data parallelism; in fact, all PEs can execute slightly different programs while still sharing the same set of control signals.
    Architectures more closely related to TCPAs are \textit{Coarse-Grained Reconfigurable Arrays}~(CGRAs)~\cite{26_CGRA_Overview,24_CGRA_Taxonomy}.
    CGRAs typically map dataflow graphs (DFGs) of loops written in C/C++ onto their PEs by assigning each operation in the DFG both the start time and the PE on which the operation is executed~\cite{TirelliSAFDAMAP24, WangTOSAAPA25, edgeCGRA, KoulMSTNZLSCMSDDCKFHNSTBDMTRBFHBHT23}.
    In contrast to TCPAs, where a global controller (GC) generates the loop control outside the array, CGRAs typically handle loop control---both for the loop itself and for conditional statements in the loop body---through predication within the array itself, rather than offloading it to an external controller.
    This makes the architecture flexible, but control overhead remains distributed across PEs, \eg, \cite{das2018efficient}.
    Some authors have also explored offloading the address generation to dedicated controllers~\cite{KoulMSTNZLSCMSDDCKFHNSTBDMTRBFHBHT23}, whereas the loop control is still done within the PEs.
    Closer to our work, \cite{BednaraHT02} propagate simple start/stop signals along the boundary to control processor arrays.
    Unlike their distributed approach, our method centralizes control in a Global Controller that evaluates arbitrary loop conditions.
\section{Control Signals}
\label{sec:control}
    In this section, control signal allocation, generation, and propagation are introduced.
    It concludes with the control instruction within each FU of a PE.

    \subsection{Control Allocation}
        TCPAs exploit parallelism not only spatially across the PE array, but also temporally through software pipelining, where multiple iterations overlap on the same PE by sharing FUs in a modulo schedule.
        Specifically, after scheduling, each equation $\plaEquation$ is assigned a start time $\plaScheduleOperationStartTime$ for every iteration $\plaIntraIteration$ in its domain ($\plaIntraIteration \in \plaIntraIterationSpace$).
        When such a scheduled time satisfies $\plaScheduleOperationStartTime > \plaScheduleInitiationInterval$, it starts after the next iteration begins, causing consecutive iterations to overlap temporally.
        Now, two consecutive iterations may require different instructions to be executed due to the condition spaces $\plaConditionSpace$ of equations within the loop description.
        This is realized by different microprograms for classes of iterations sharing the same instructions.
        However, if their execution overlaps, the PEs must either support multiple parallel program contexts, or, the overlaps must be resolved at compile time.
        To keep hardware simple, we choose the second option introduced next.
        We resolve all potential overlaps of iterations, by generating a set of microprograms that contains any possible execution sequence that can occur due to overlapping iterations.
        This can be done by adjusting the condition spaces after scheduling.
        Whenever an equation $\plaEquation$ is assigned a start time $\plaScheduleOperationStartTime > \plaScheduleInitiationInterval$, its condition space $\plaConditionSpace$ is shifted
        $\lfloor \frac{\plaScheduleOperationStartTime}{\plaScheduleInitiationInterval} \rfloor$ iterations in time, meaning that we must find for every iteration $\plaIntraIteration \in \plaIntraIterationSpace$, a corresponding iteration $\plaIntraIteration'$ that starts $\lfloor \frac{\plaScheduleOperationStartTime}{\plaScheduleInitiationInterval} \rfloor \cdot \plaScheduleInitiationInterval$ cycles later.
        Formally:
        $$
        \plaScheduleOperationShiftDomainDefinition
        $$
        This transformation is not linear as it deforms condition spaces from single polyhedra into more general unions~(or lists) of polyhedra.
        Note that this often even requires an increase in the iteration space in form of an epilog domain to complete all running iterations.
        Finally, the operation’s start time is adjusted such that $\plaScheduleOperationStartTime < \plaScheduleInitiationInterval$:
        $$
        \plaScheduleOperationWrap
        $$
        \begin{figure*}
            \resizebox{\textwidth}{!}{
            \includegraphics[width=\textwidth]{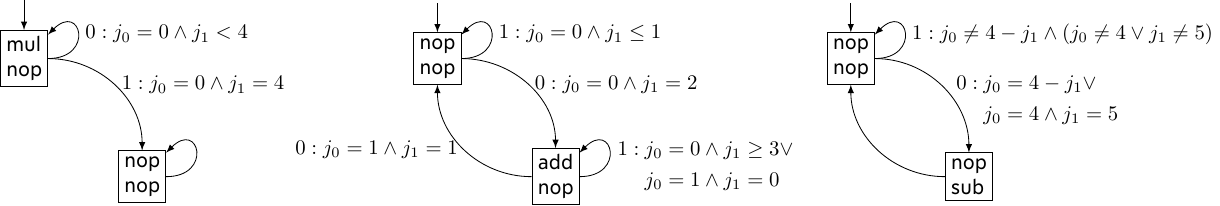}
            }
            \caption{%
                Example of 3 control graphs.
                Each node represents a program block~(sequence of instructions) and each edge denotes a branch between such program blocks. The annotated conditions indicate when the corresponding branch must be taken.}
            \label{fig:ctrlGraphs}
        \end{figure*}
        \begin{figure*}
            \resizebox{\textwidth}{!}{
            \includegraphics[width=\textwidth]{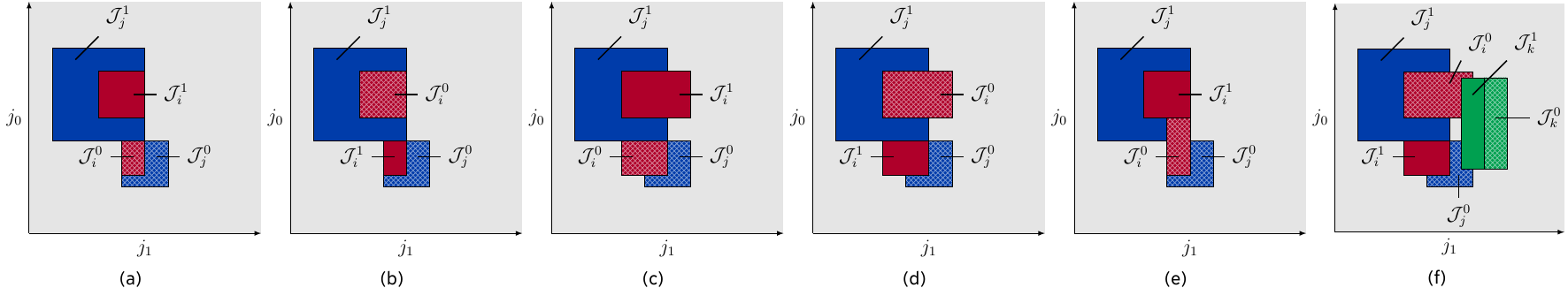}
            }
            \caption{%
                A two-dimensional iteration space with different control conditions~(red, blue, green) is shown for 6 different cases.
                The solid rectangles denote the one domain, and the crosshatched rectangle denotes the corresponding zero domain.
                In (a)--(d), both control conditions can be expressed by a single control signal, while in (e), both control conditions are conflicting.
                Although the three control conditions in (f) are pairwise compatible, they cannot be unified into a single control condition.
            }
            \label{fig:ctrlGroups}
        \end{figure*}
        This preserves the original schedule while modifying only the iteration domains to ensure non-overlapping execution.
        Thus, all operations within any iteration are scheduled within $\plaScheduleInitiationInterval$ cycles.
%
        Next, since TCPAs use OIP, we identify for every operation the assigned FU and construct for each FU a set of branch-free programs called program blocks, \ie, a sequence of not more than $\plaScheduleInitiationInterval$ instructions.
        The order in which program blocks are executed by the FU is given by a control graph~(see \figRef{fig:ctrlGraphs}), where an edge from node $\plaNode_i$ to node $\plaNode_j$ indicates program block $\plaNode_j$ must execute after $\plaNode_i$.
        If there are multiple outgoing edges, each edge is also annotated with a condition, \ie, when to take that edge.
        The conditions are computed from the iteration domains, so they depend statically on the loop indices.
        Typically, the control graph is transformed so that each node has an outdegree of at most two.
%
        We formally denote the condition of a node $\plaNode_i$ within the control graph as a binary control condition $\plaCtrlCondition_i$:
        $$\plaCtrlConditionDefinition$$
        $\plaCtrlConditionZeroDomain_i$ is called the zero domain of $\plaCtrlCondition_i$, indicating that the condition is zero in these iterations, while the so-called one domain $\plaCtrlConditionOneDomain_i$ of $\plaCtrlCondition_i$ requires that the control condition is one in the contained iterations.
        This binary condition identifies which of the two successor nodes of $\plaNode_i$ must be executed next.
        The assignment between the logical values (0 or 1) and the successor node is called the polarity of the control condition $\plaCtrlCondition_i$.
        With a positive polarity, the zero domain $\plaCtrlConditionZeroDomain_i$ denotes the part of the iteration space in which, after executing the program block of $\plaNode_i$, the execution must continue with the first of the two successor nodes of $\plaNode_i$.
        Analogously, the one domain $\plaCtrlConditionOneDomain_i$ refers to the part of the iteration space in which the second edge must be taken.
        On the other hand, with a negative polarity, the assignment is reversed, \ie, $\plaCtrlConditionZeroDomain_i$ corresponds to the second successor, and $\plaCtrlConditionOneDomain_i$ corresponds to the first successor.
        Thus, by changing the polarity, the zero and one domains are swapped.
        Since the order of successors is arbitrary, the polarity can be freely chosen or swapped without changing the semantics of the control flow.
%
        Control graphs and associated conditions are generated separately for each tile/PE.
        However, all control conditions of all tiles are collected in a single set $\plaCtrlConditionSpace$, so a single GC can evaluate all conditions.
        Since there are typically hundreds of such control conditions, the set $\plaCtrlConditionSpace$ must be simplified to reduce hardware overhead.
        This is achieved by two transformations introduced in the following.
        Consider \figRef{fig:ctrlGroups}.
        Each subplot shows a two-dimensional iteration space with different control conditions highlighted.
        Shown as the solid rectangles are the one domains of the control conditions $\plaCtrlCondition_i$~(red), $\plaCtrlCondition_j$~(blue), and $\plaCtrlCondition_k$~(green), with the zero domains being shown as the crosshatched rectangles.
        In the following, each of the 6 cases is introduced.
        \headingBlock{\textbf{(a)--(b)}:}
        As seen in (a), the blue condition $\plaCtrlCondition_j = (\plaCtrlConditionZeroDomain_j, \plaCtrlConditionOneDomain_j) $ fully encapsulates the red condition $\plaCtrlCondition_i = (\plaCtrlConditionZeroDomain_i, \plaCtrlConditionOneDomain_i)$---for all iterations in which $\plaCtrlCondition_i$ is zero, $\plaCtrlCondition_j$ is also zero, and vice versa.
        Hence, evaluating $\plaCtrlCondition_j$ also yields $\plaCtrlCondition_i$.
        The latter is thus obsolete and must not be further considered.
        In (b), we observe that the one domain of $\plaCtrlCondition_j$~(solid blue) encapsulates the zero domain of $\plaCtrlCondition_i$~(crosshatched red), while the zero domain of $\plaCtrlCondition_j$~(crosshatched blue) encapsulates the one domain of $\plaCtrlCondition_i$~(solid red).
        In this case, we can change the polarity of $\plaCtrlCondition_i$ as explained earlier.
        This swaps the zero and one domains of $\plaCtrlCondition_i$ such that it is fully encapsulated by $\plaCtrlCondition_j$.
        We denote both cases formally as:
        $$\plaCtrlPrime \lor \plaCtrlInversePrime$$
        By filtering out all obsolete control conditions, we obtain the set of so-called prime control conditions $\plaCtrlPrimeConditionSpace$.
        Finding the minimal set of prime control conditions is achieved by an exhaustive search described as \algoRef{algo:ctrlPrime} with complexity $\mComplexity{\mquant{\plaCtrlConditionSpace}^2}$.
        Given all control conditions~$\plaCtrlConditionSpace$, it iterates over all $\plaCtrlCondition_i$ and known prime conditions~$\plaCtrlCondition_j$.
        In the first if condition, it is tested whether $\plaCtrlCondition_j$ encapsulates $\plaCtrlCondition_i$.
        If this is the case, $\plaCtrlCondition_i$ is not prime, and we continue with the next unchecked control condition.
        However, it is also possible that $\plaCtrlCondition_i$ encapsulates $\plaCtrlCondition_j$, which is checked in the second if condition.
        In this case, $\plaCtrlCondition_j$ is not prime and removed from $\plaCtrlPrimeConditionSpace$ before inserting $\plaCtrlCondition_i$ as a potential prime control condition.
        If none of the if conditions is true, $\plaCtrlCondition_i$ is added to $\plaCtrlPrimeConditionSpace$.
        After this algorithm, any remaining prime control conditions cannot be fully represented by any other control conditions.
        However, we can further reduce the number of control conditions by combining compatible control conditions into unified control conditions as shown next.
        \tiny
        \plaCtrlConditionPrimeAlgo
        \normalsize
%
        \headingBlock{\textbf{(c)--(d)}:}
        Consider again \figRef{fig:ctrlGroups}.
        In (c), the domains of $\plaCtrlCondition_i$ and $\plaCtrlCondition_j$ overlap, but unlike (a) and (b), $\plaCtrlCondition_j$ does not fully encapsulate $\plaCtrlCondition_i$.
        However, we can still unify both into a new condition as long as their zero and one domains do not contradict each other, \ie, for any iteration where $\plaCtrlCondition_i$ is zero, $\plaCtrlCondition_j$ must not be one, and vice versa.
        Similar to (b), in (d), both control signals can only be unified if the polarity of $\plaCtrlCondition_i$ is reversed.
        This again results in a swapped zero and one domain of $\plaCtrlCondition_i$.
        Thus, formally, two control conditions $\plaCtrlCondition_i$ and $\plaCtrlCondition_j$ can be unified if
        $$\plaCtrlCompability \lor \plaCtrlInverseCompability$$ The unified control condition $\plaCtrlCondition'$ is then given as
        $$\plaCtrlCondition' = \plaCtrlConditionAdd ,\quad \plaCtrlCondition' = \plaCtrlConditionInverseAdd\quad\text{resp.}$$
        If we apply this repeatedly to the set of control conditions, we obtain the set of unified control conditions $\plaCtrlConditionUnified$, which cannot be further unified.
        Finding an optimal, \ie, the smallest possible, set of unified control conditions is, however, non-trivial.
        \headingBlock{\textbf{(e)--(d)}:}
        Consider \figRef{fig:ctrlGroups}~(e).
        Both control conditions conflict with each other as the zero domain of $\plaCtrlCondition_i$~(crosshatched red) overlaps both the zero and one domain of $\plaCtrlCondition_j$~(blue).
        However, such conflicts are often not as trivial.
        Now, consider (f).
        It extends \figRef{fig:ctrlGroups}~(d) by an additional control condition $\plaCtrlCondition_k$, shown in green.
        As explained earlier, $\plaCtrlCondition_i$~(shown in red) and $\plaCtrlCondition_j$~(shown in blue) can be unified by switching the polarity of $\plaCtrlCondition_i$.
        Also, both $\plaCtrlCondition_i$ and $\plaCtrlCondition_j$ can be unified individually with $\plaCtrlCondition_k$~(shown in green).
        However, although all three control conditions are pairwise compatible, \ie, they can be unified, we cannot unify them into a single control condition.
        As can be seen in the figure, $\plaCtrlCondition_i$ and $\plaCtrlCondition_j$ can only be unified by changing the polarity of one of them, \ie, by swapping the one domain~(solid color) and the zero domain~(crosshatched color).
        But $\plaCtrlCondition_k$ overlaps the zero domain of both $\plaCtrlCondition_i$ and $\plaCtrlCondition_j$.
        Thus, $\plaCtrlCondition_i$ and $\plaCtrlCondition_j$ can only be unified with $\plaCtrlCondition_k$ if they have the same polarity.
        This contradicts the fact that $\plaCtrlCondition_i$ and $\plaCtrlCondition_j$ must have reversed polarity to be unified.
        Therefore, there is no solution.
        This can be viewed as a min-clique problem on a compatibility graph, where each node represents a control condition and each edge represents compatibility of two conditions with respect to their unification.
        However, due to polarity, the edges of this compatibility graph are conditional, \ie, an edge may be present only if another edge is absent.
        This makes finding an optimal solution infeasible for practical problem sizes.
        Consequently, we use a greedy approach as shown in \algoRef{algo:ctrlUnified} with complexity $\mComplexity{\mquant{\plaCtrlConditionSpace}^2}$.
        It iterates in random order over all prime control conditions $\plaCtrlCondition_i$ and known unified control conditions $\plaCtrlCondition_j$.
        It checks in the first if-condition whether $\plaCtrlCondition_i$ and $\plaCtrlCondition_j$ can be unified, and checks in the second if-condition whether changing the polarity of $\plaCtrlCondition_i$ makes it compatible with $\plaCtrlCondition_j$.
        In each case, $\plaCtrlCondition_i$ and $\plaCtrlCondition_j$ are unified and added to the set of known unified control conditions $\plaCtrlConditionUnified$.
        Otherwise, if no compatible control condition $\plaCtrlCondition_j$ could be found for $\plaCtrlCondition_i$, $\plaCtrlCondition_i$ is added directly to $\plaCtrlConditionUnified$.
        The algorithm performs differently on different random orders, hence it is called repeatedly with varying orders to find a smaller solution.
        In practice, we found that this converges quickly as the final result is typically found already within the first few tries.
        In this work, we take the smallest solution after 100 tries, which determines the final set of unified control conditions $\plaCtrlConditionUnified$, each requiring one physical control signal.
        Note that in this section, we have omitted the bookkeeping of the polarity of all original control conditions $\plaCtrlCondition_i \in \plaCtrlConditionSpace$.
        \plaCtrlConditionUnifiedAlgo

    \subsection{Control Generation}
    \label{sec:ctrl:gen}
        Control signals are computed by a dedicated global controller.
        Each control signal is a binary signal that is one in all iterations $\plaIntraIteration$ belonging to the one domain $\plaCtrlConditionOneDomain_i$ of the assigned $\plaCtrlCondition_i$, and zero otherwise.
        Thus, the GC must evaluate all $\plaCtrlConditionOneDomain_i$ at runtime for every iteration.
        Note that any domain $\plaCtrlConditionOneDomain_i$ can be decomposed into a union of polyhedra, which can be viewed as a disjunction of conjunctions of literals.
        The literals can be further categorized into one of the following types:
        \begin{itemize}
            \item \textbf{Constant equality bound}: A comparison of a loop index with a constant value for equality, \eg, $j_0 = 3$.
            \item \textbf{Constant lower bound}: A comparison of a loop index with a constant lower bound, \eg, $j_0 \ge 3$.
            \item \textbf{Constant upper bound}: A comparison of a loop index with a constant upper bound, \eg, $j_0 \le 3$.
            \item \textbf{Affine equality bound}: An affine expression of loop indices that is evaluated for equality, \eg, $j_0 = j_1 - 1$.
            \item \textbf{Affine inequality bound}: An affine expression of loop indices that is evaluated for inequality, \eg, $j_0 \ge j_1 - 1$.
        \end{itemize}
        \begin{figure}
            \centering
            \resizebox{0.5\textwidth}{!}{
            \includegraphics[width=0.5\textwidth] {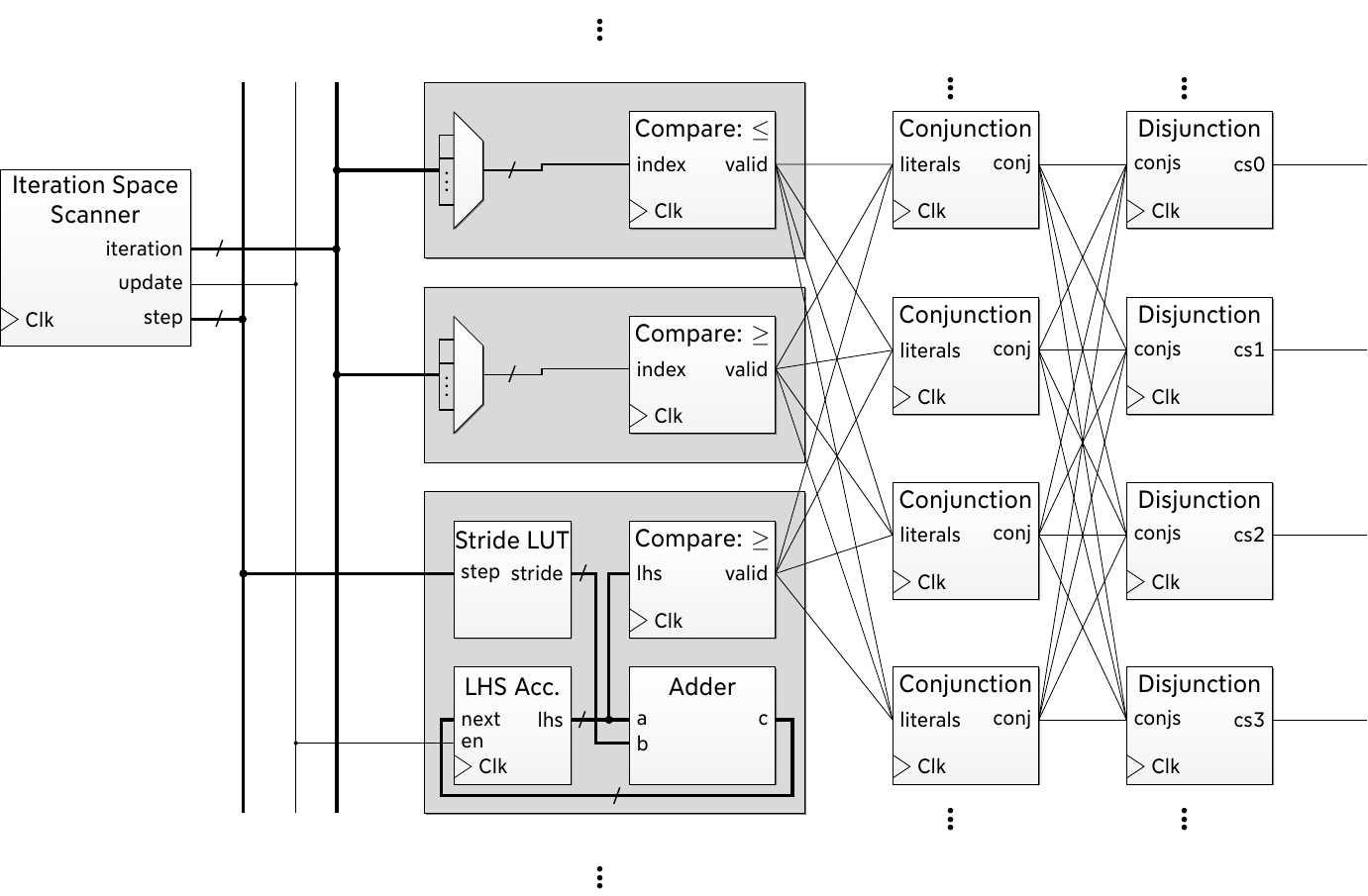}
            }
            \caption{%
                Architecture of the Global Controller (GC).
                It consists of an Iteration Space Scanner~(left), which feeds into a number of evaluators~(center), whose outputs are combined into conjunctions and disjunctions~(right).
            }
            \label{fig:gc_arch}
        \end{figure}
        Now consider \figRef{fig:gc_arch}, showing the simplified GC architecture~(control logic omitted for clarity).
        On the left is an \texttt{Iteration Space Scanner}.
        It must be configured with the iteration space $\plaIntraIterationSpace$ and the initiation interval $\plaScheduleInitiationInterval$ before computing the current \texttt{iteration} representing $\plaIntraIteration$ every $\plaScheduleInitiationInterval$ cycles.
        Without loss of generality, all iteration spaces can be transformed so that the first scheduled iteration is always $\plaIntraIteration = (0,\dots, 0)\mtrans$, where $\plaIntraIterationElement_0$ is scanned first.
        This dramatically simplifies the hardware for generating the current iteration.
        Furthermore, \texttt{update} indicates that a new iteration will be computed in the next cycle and \texttt{step} indicates the dimension that will be incremented next within the iteration vector.
        The \texttt{Iteration Space Scanner} feeds an array of evaluators shown in the center~(gray boxes), where each box denotes a different type of evaluator:
        \begin{itemize}
            \item \textbf{Lower bound evaluator}: An evaluator supporting constant equality and constant lower bounds.
            \item \textbf{Upper bound evaluator}: An evaluator supporting constant equality and constant upper bounds.
            \item \textbf{Affine bound evaluator}: An evaluator supporting affine equality and affine inequality bounds.
        \end{itemize}
        Each literal that occurs within a one domain of a control signal is mapped to an evaluator that supports that literal type.
        Therefore, a typical GC requires dozens of evaluators of each type.
        \figRef{fig:gc_arch} shows, at the top, a lower-bound evaluator containing a configurable multiplexer that selects an index from the current iteration and passes it to a comparator that compares the selected index to a configured constant value.
        The comparator can be configured to use either $\leq$ or $=$ for comparison, \ie, it can evaluate both constant lower bounds and constant equality bounds.
        Similarly, for the upper bound evaluator shown in the center.
        The last evaluator type, \ie, the affine bound evaluator, is shown at the bottom.
        Evaluating any affine condition requires the computation of $u = \mvec{a} \cdot \plaIntraIteration$, where $\mvec{a} \in \mbaseSet{Z}^\plaDimension$ is an arbitrary weight vector.
        $u$ can then be compared against a constant in order to implement the respective affine condition.
        To avoid an expensive scalar product in hardware, the affine bound evaluators use a stride-based approach introduced below.
        Let $u$ be known for an iteration $\plaIntraIteration$, and consider the following computation of $u'$ of the successor iteration $\plaIntraIteration'$:
        $$u' = \mvec{a} \cdot \plaIntraIteration' = \mvec{a} \cdot (\plaIntraIteration + \epsilon) = u + \mvec{a} \cdot \epsilon$$
        $\epsilon$ denotes the change between consecutive iterations; since the iteration space is traversed step by step, there are exactly $\plaDimension$ different $\epsilon$ in an $\plaDimension$-dimensional loop.
        Therefore, all possible strides $\mvec{a} \cdot \epsilon$ can be precomputed statically and stored in hardware.
        Thus, the scalar product reduces to a simple left-hand-side accumulator~(\texttt{LHS Acc.}) and stride look-up-table~(\texttt{Stride LUT}).
        Since we consider only normalized iteration spaces, \ie, the first iteration is always zero, the initial value of the accumulator is also zero.
        The \texttt{step} output of the iteration space scanner indicates which stride the \texttt{Adder} should add onto the current accumulator value~(\texttt{lhs}).
        With \texttt{update}, the hardware signals that the accumulator should be updated so that for each scanned iteration, the accumulated left side of the affine condition is correctly computed.
        Finally, the current accumulator value is, similar to the upper bound evaluator type, compared against a constant supporting both $\geq$ and $=$.
        Each evaluator returns \texttt{valid}, a binary signal indicating whether its assigned literal is true or not.
        These valid signals are propagated to \texttt{Conjunctions}~(AND) and \texttt{Disjunctions}~(OR), computing any control condition as a disjunction of conjunctions.
        Conjunctions and disjunctions are configured with masks indicating which inputs are ANDed or ORed.
        Note that the number of evaluators, conjunctions, and disjunctions is not fixed in all GCs and is part of the architectural parameters.
        Finally, each disjunction evaluates exactly one control signal that is passed to the PEs.
    \subsection{Control Propagation}
        \begin{figure}
            \centering
            \resizebox{0.3\textwidth}{!}{
            \includegraphics[width=0.5\textwidth] {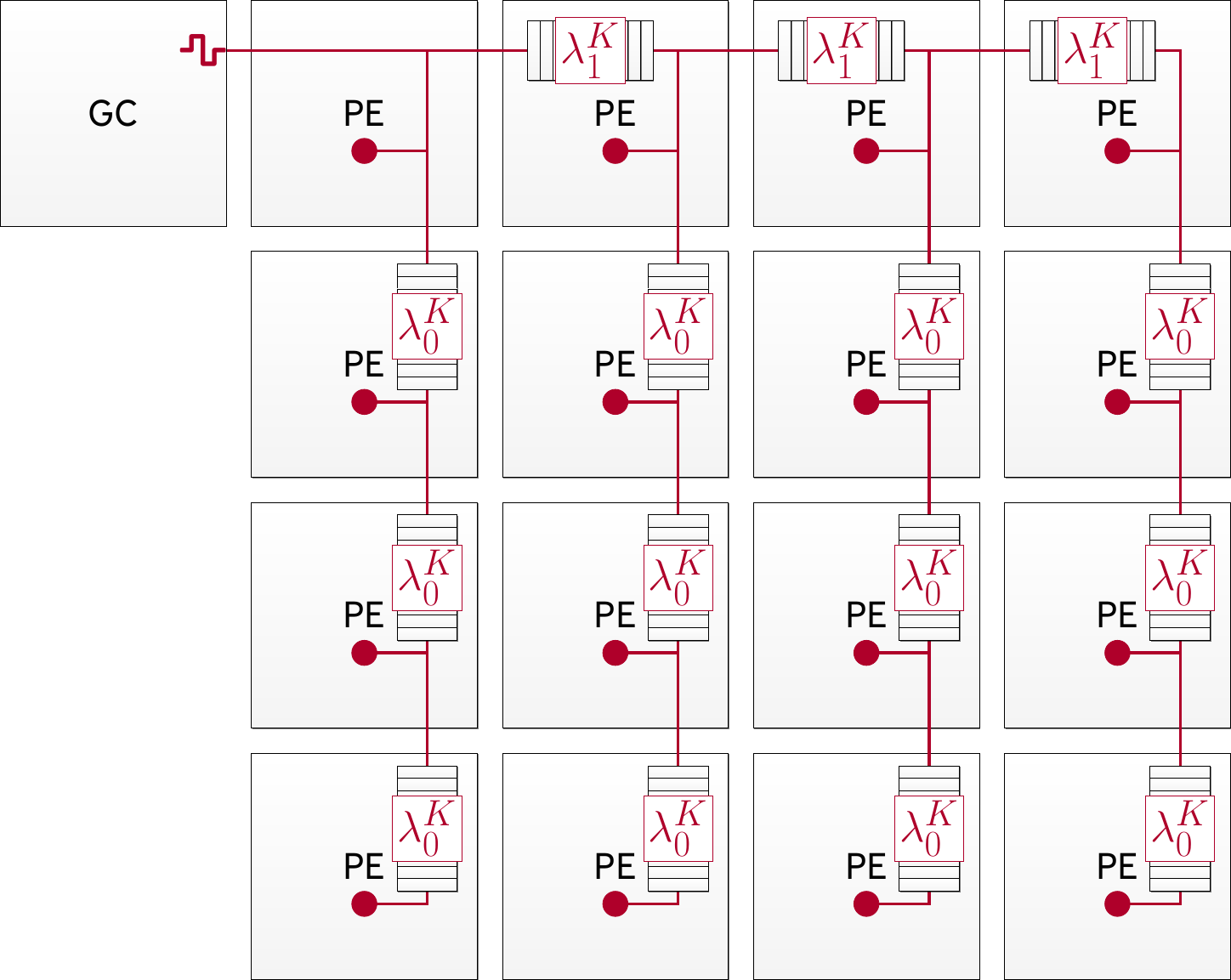}
            }
            \caption{%
                Control network in a $4 \times 4$ TCPA.
                Control signals are generated in the GC and propagated to all PEs.
                Delay elements within each PE shift the incoming signals in time so that they arrive according to the loop schedule.
            }
            \label{fig:ctrl_network}
        \end{figure}
        Control signals are propagated from the GC via a dedicated network~(\figRef{fig:ctrl_network}).
        The same control signals are shared by all PEs, but iteration $\plaIntraIteration$ starts $\plaTiledInterScheduleVectorElement_0$ cycles later on a PE than its top neighbor and $\plaTiledInterScheduleVectorElement_1$ cycles later than its left neighbor.
        Each PE delays incoming signals by a configurable number of cycles, implemented as a shift register or, for large latencies, a timestamp FIFO encoding each signal transition as an integer.
        Trade-offs between the two approaches are evaluated in \secRef{sec:eval}.
    \subsection{Control Instruction}
\begin{figure}
            \centering
            \resizebox{0.5\textwidth}{!}{
            \includegraphics[width=0.5\textwidth] {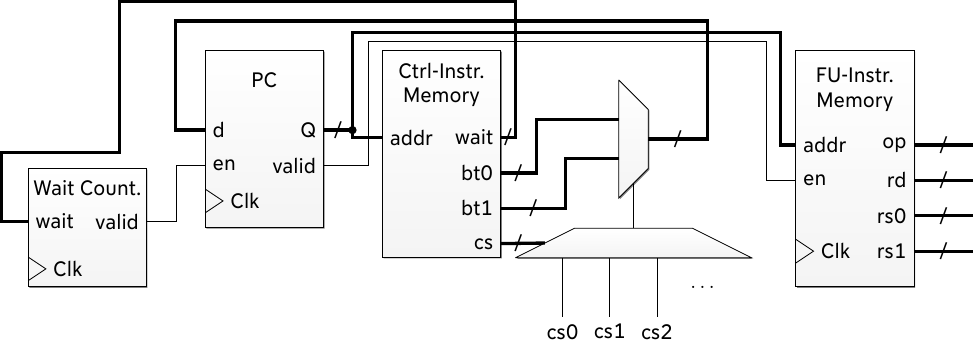}
            }
            \caption{Architecture of the instruction sequencer within each functional unit of a PE. The shown control signals~(cs0, cs1, cs2, \dots) originate directly from the control network shown in \figRef{fig:ctrl_network}.}
            \label{fig:ctrl_pipe}
        \end{figure}

        Consider a microprogram of a FU.
        Each instruction encodes an operation~\texttt{op} on some registers~\texttt{rd}, \texttt{rs0}, \texttt{rs1} as:
        \begin{align*}%
              &\texttt{op}~~\texttt{rd}~~\texttt{rs0}~~\texttt{rs1}
        \end{align*}%
        The microprogram is stored in an instruction memory within each FU, such that the first program block to execute begins at address 0.
        The program counter is initially zero, incremented with each instruction, and at the last instruction of a program block, set to the start address of the next block.
        The update of the program counter is encoded in control instructions of form:
        \begin{align*}
          &\texttt{bt0}~~\texttt{bt1}~~\texttt{cs}~~\texttt{wait}
        \end{align*}
        If the control signal \texttt{cs} is active, \ie, the respective control condition is valid in the current iteration, the program counter is set to the branch target \texttt{bt0}, or, otherwise, to \texttt{bt1}.
        In the last instruction of a program block, branch targets are set to the start addresses of the next program blocks; in all other instructions both targets are the same, implementing a simple unconditional jump.
        Such a jump is also used when there is only one possible successor to the current program block.
        Program blocks often contain long nop sequences that fill the small memory available.
        To avoid such overhead, each control instruction has an additional \texttt{wait} field, indicating that the FU must wait a specified number of cycles before starting the next instruction.
        Now, consider \figRef{fig:ctrl_pipe}.
        It shows the simplified instruction sequencer within each PE that implements the control instructions.
        The current program counter~(\texttt{PC}) is used to address an asynchronous memory containing the control instructions~(\texttt{Ctrl-Instr. Memory}).
        The \texttt{cs} field selects the corresponding control signal, which is used immediately to select one of the two branch targets \texttt{bt0} and \texttt{bt1}.
        Additionally, a counter~(\texttt{Wait Count.}) receives the \texttt{wait} part of the instruction.
        The counter triggers its output signal~(\texttt{valid}) after \texttt{wait} cycles indicating that the program counter can now be updated with the selected branch target.
        Data dependences are resolved at compile time during scheduling, so no reservation stage is needed.
        Since the FU must be able to issue a new instruction every single cycle~(\texttt{wait} = 0), the PC update must also be combinatorial.
        Finally, the current PC is also routed to a second synchronous instruction memory containing the FU instructions~(\texttt{FU-Instr. Memory}).
        This separation reduces the size of expensive asynchronous memory within a FU.
        The read FU instructions are then passed to an instruction decoder, the register file, and finally the FU itself.
        Note that such later stages are not connected to the instruction sequencer, avoiding all pipeline hazards.
\section{Evaluation}
\label{sec:eval}
    We evaluate the control allocation~(number of required signals), the hardware cost for signal generation and propagation, and the overhead within each FU, and show that our approach is able to massively reduce the number of control signals while keeping the required hardware overhead to a minimum. 
    \begin{table*}
        \caption{Control required for different benchmarks}
        \label{table:bench}
        \centering
        \resizebox{1.0\textwidth}{!}{
        \includegraphics[width=0.5\textwidth] {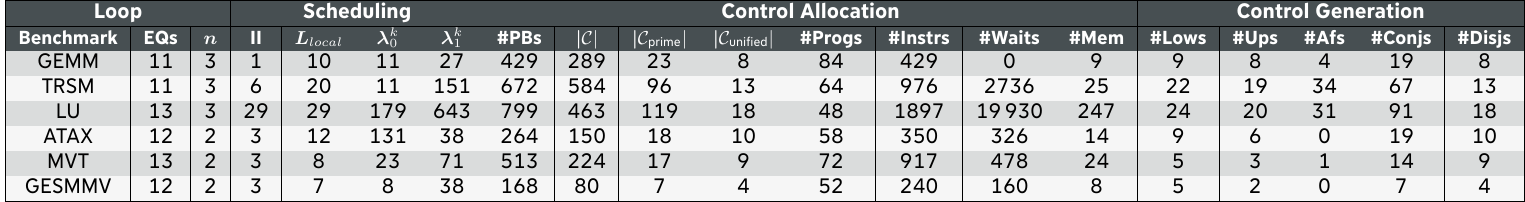}
        }
    \end{table*}
    \subsection{Control Allocation}
        \tableRef{table:bench} presents control flow requirements for six PolyBench kernels~\cite{polybench}: GEMM~(matrix-matrix multiplication), TRSM~(triangular solve), LU~(LU decomposition), and ATAX, MVT, and GESMMV~(fused matrix-vector products).
        Each was implemented as an $\plaDimension$-dimensional PRA and scheduled for $n=20$ on a $4 \times 4$ TCPA.
        GEMM achieves an optimal $\plaScheduleInitiationInterval = 1$ with up to 10 overlapping iterations, while LU’s long iterations~($\plaScheduleLocalLatency = 29$) prevent any pipelining.
        Most benchmarks require shift registers of up to 151 cycles to delay control signals, which can be implemented with configurable registers of that length.
        LU, however, is an outlier, requiring a significantly longer shift register of 643 cycles.
        In such cases, timestamp-based FIFOs may offer advantages over simple shift registers.
        The table also summarizes the results of control allocation.
        The schedule yields a set of program blocks~(\#PBs) and a corresponding control condition space~$\plaCtrlConditionSpace$.
        By applying our reduction techniques~($\plaCtrlPrimeConditionSpace, \plaCtrlConditionUnified$), we report for each benchmark the number of required functional unit (FU) programs~(\#Progs), the total instruction count~(\#Instrs), the number of skipped nops due to wait counters~(\#Waits), and the size of the longest program~(\#Mem) that must fit in an FU’s instruction memory.
        The reduction to prime control conditions alone achieves a substantial decrease in control condition count by a factor of $\factorRange{4}{13}$, while combining into unified conditions provides an additional reduction of $\factorRange{2}{7}$.
        Benchmarks with more prime conditions, such as LU~($\factor{4}$ for prime), benefit more from unification~($\factor{7}$), whereas those with fewer prime conditions, like GEMM~($\factor{13}$), see less unification potential~($\factor{3}$).
        Together, these reductions yield an overall decrease in control conditions by a factor of $\factorRange{15}{45}$.
        Eliminating nops via wait counters is another critical optimization.
        It reduces instruction memory requirements by up to $\qty{91}{\percent}$ in LU and still achieves a $\qty{34}{\percent}$ reduction in simpler benchmarks like MVT.
        Notably, when $\plaScheduleInitiationInterval = 1$, program blocks are of unit length and require no nop padding.
        Hence, benchmarks like GEMM do not require wait cycles.
        Finally, generating control signals for FU programs requires the GC to evaluate disjunctions~(\#Disjs) of conjunctions~(\#Conjs) of lower-bound~(\#Lows), upper-bound~(\#Ups), and affine-bound evaluators~(\#Afs).
        Control-intensive benchmarks with triangular condition spaces, such as TRSM and LU, require significantly more evaluators, especially affine ones, as well as more conjunctions and disjunctions.
        Conversely, the other, simpler benchmarks require fewer hardware resources.
        Across all benchmarks, the total compilation time for deriving and allocating all control signals was less than one minute on a standard Intel Core i7 machine.

    \subsection{Control Generation}
\begin{figure*}
        \centering
        \resizebox{1.0\textwidth}{!}{
        \includegraphics[width=0.5\textwidth] {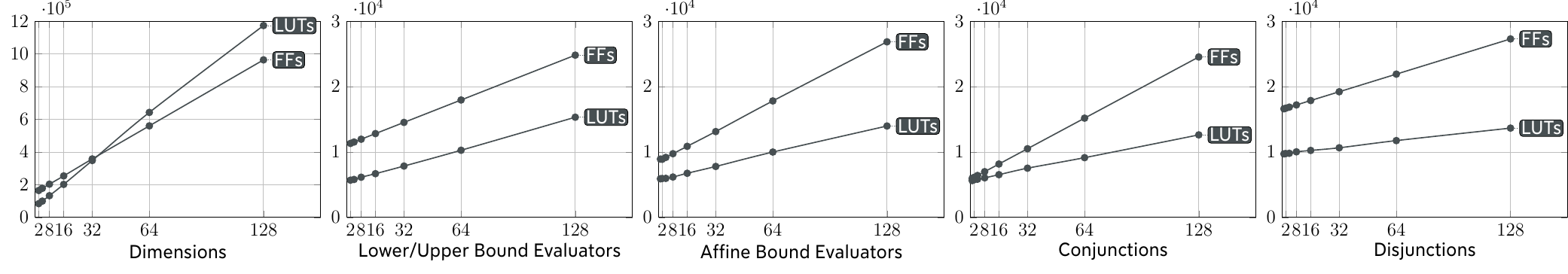}
        }
        \caption{The required LUTs and FFs for the GC when scaling individual hardware parameters.}
        \label{fig:gc_cost}
    \end{figure*}

        To evaluate the hardware overhead for generating the control signals, we synthesized a VHDL description of a $4 \times 4$ TCPA capable of executing the loops shown in \tableRef{table:bench} for an AMD/Xilinx Ultrascale+ FPGA target using Vivado.
        This design achieved a clock frequency of $\qty{250}{\mega\hertz}$ and required a total of $\qty{234331}{\luts}$ and $\qty{222209}{\ffs}$.
        However, only a small fraction of these resources were used for the GC, which required only $\qty{9764}{\luts}$~($\qty{4.17}{\percent}$) and $\qty{17860}{\ffs}$~($\qty{8.04}{\percent}$), roughly the cost of a single PE for comparison.
        In this implementation, the GC supports 4 dimensions, 32 lower-bound evaluators, 32 upper-bound evaluators, 65 affine-bound evaluators, 83 conjunctions, and 18 disjunctions, \ie, control signals.
        To further investigate the scalability of the GC, we varied these hardware parameters and synthesized the GC alone.
        The resulting costs are shown in~\figRef{fig:gc_cost}.
        The x-axis indicates the hardware parameter that is being changed, while the other parameters are constant.
        We observe that, in all subplots, the costs scale linearly but at different rates.
        Most critical is obviously the number of dimensions that lead to a linear increase in FF and a super-linear scaling of LUTs.
        More dimensions require a larger iteration space scanner, larger multiplexers in the lower and upper bound evaluators, and larger stride-tables in the affine bound evaluators.
        When pushing this parameter to the extreme~(128 loop dimensions), the GC still requires only about half of the FPGA, or, 12 PEs, which also indicates the only design point in which the target frequency could not be achieved.
        For realistic loops with $2-8$ dimensions, less than 20k LUTs and FFs are sufficient.
        The other 4 subplots show that changing the number of evaluators, conjunctions, and disjunctions, increases the hardware cost only sub-linearly.
        Here, the number of conjunctions scales the worst~(steepest increase), followed by the number of affine-bound evaluators, lower/upper bound evaluators, and, finally, the number of disjunctions.
        Notably, the GC requires more FFs than LUTs, \ie, it consists mostly of configuration registers.
        This is especially critical in the affine bound evaluators that require each a stride-table, whereas the lower- and upper-bound evaluators only store each the index of the selected loop index.
        As a result, such simple evaluators require fewer FFs but slightly more LUTs.
        Note also that each evaluator, regardless of its type, also increases the cost of each conjunction for the additional input.
        For example, with 83 conjunctions, each evaluator attributes 83 bits of configuration data to the conjunction masks.
        Finally, the number of disjunctions is, for the GC, mostly irrelevant.
        Since reasonable TCPA implementations typically require no more than 32 control signals~(see \tableRef{table:bench}), the increase in hardware resources within the GC is negligible.
    \subsection{Control Propagation}
        \begin{figure}
            \centering
            \resizebox{0.5\textwidth}{!}{
            \includegraphics[width=0.5\textwidth] {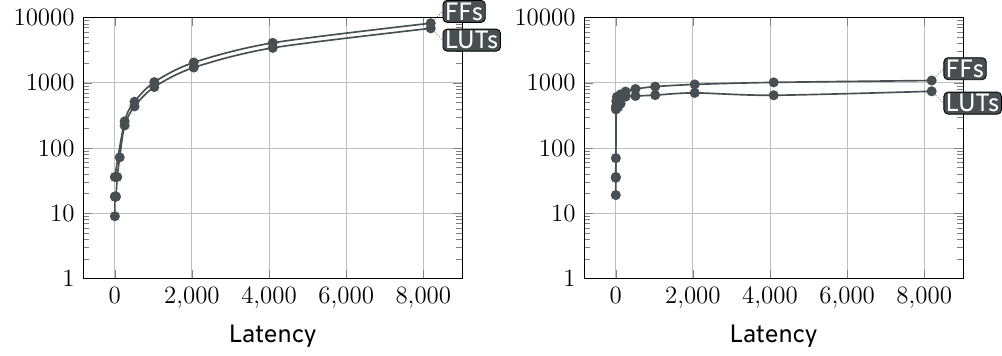}
            }
            \caption{The LUTs and FFs required for a shift register~(left) and a timestamp FIFO~(right) when scaling the maximum latency of the 18 control signals.}
            \label{fig:ic_cost}
        \end{figure}

        Every PE must delay its incoming control signals by a configurable latency, \ie, a number of cycles.
        As mentioned earlier, there exist two different approaches for such a hardware unit---configurable shift-registers and timestamp FIFOs.
        \figRef{fig:ic_cost} compares shift registers and timestamp FIFOs for different latencies.
        Shift registers scale linearly; timestamp FIFOs converge near $\num{1000}$ LUTs and FFs by using RAMB18/RAMB36 blocks~(1 RAMB18 up to 1024 entries, 1 RAMB36 up to 2048, etc.).
        For fewer than 1024 entries, shift registers are cheaper, but for larger latencies, timestamp FIFOs are preferable in LUT-constrained designs.
        This significant overhead for these delay units is also the reason why it is essential to optimize the number of required control signals as much as possible.
        The resources spent on the GC can be easily amortized since it exists only once, but an 8192-entry long $\qty{18}{\bit}$-shift register costs almost as much as the GC and has to be instantiated in every single PE.
        Finally, note that in the $4 \times 4$ TCPA design mentioned earlier, we used timestamp FIFOs that support a latency of up to 4096, which still only accounts for $\qty{4}{\percent}$ of LUTs, $\qty{5}{\percent}$ of FFs, and $\qty{60}{\percent}$ of BRAMs of the entire TCPA design.
        As can be seen in \tableRef{table:bench}, most benchmarks, however, do not even require such long latencies.

    \subsection{Control Instruction}
        The instruction sequencer of a single FU, shown in \figRef{fig:ctrl_pipe}, uses only about $\qty{0.5}{\percent}$ of the resources within an entire PE.
        The asynchronous instruction memory for the control instructions requires, depending on its size $16-\qty{64}{\luts}$ as LUTRAM with a total logic overhead of only $25-\qty{44}{\luts}$ and $23-\qty{29}{\ffs}$.
        The synchronous data instruction memory is implemented by a single BRAM18 instance.
\section{Summary}
    In this paper, we presented a methodology to offload loop control in Tightly Coupled Processor Arrays~(TCPAs) to a single Global Controller~(GC) that generates control signals propagated through the array to all PEs in a timely manner.
    To map a given loop and schedule onto this architecture, condition spaces within the iteration space must first be adjusted so that iteration execution does not overlap.
    Each branch between two consecutive program blocks of an FU is represented by a control condition indicating which branch target to use.
    Since the number of such conditions is large, we propose finding prime control conditions and then unifying them into an even smaller set, yielding a reduction factor of $\factorRange{15}{45}$ across our benchmarks.
    Every remaining condition is implemented by a physical control signal set to one whenever the assigned condition, a disjunction of conjunctions of inequalities, is satisfied.
    Evaluation of these conditions is performed within the GC, whose architecture generates configurable control signals at low hardware cost~(less than a single PE, $<\qty{5}{\percent}$ LUTs of an entire TCPA).
    The generated signals are propagated through a control network to all PEs, where each PE delays incoming signals by a configurable number of cycles.
    Here, we introduce two different approaches.
    Shift registers offer linear hardware cost, while dedicated timestamp FIFOs achieve nearly constant LUT/FF count via BRAM usage, enabling large latencies~($>1024$) at low cost~($\qty{4}{\percent}$ of overall LUTs).
    Within PEs, control signals are evaluated by control instructions that update the PC of a tiny microprogram every cycle.
    As a result, the control flow of TCPAs requires zero runtime overhead with only moderate additional hardware cost~($<\qty{10}{\percent}$ LUTs).
\printbibliography

\end{document}

%% file: include.tex
\usepackage{bm}
\usepackage{mathrsfs}
\usepackage{xspace}
\usepackage{amsmath}
\usepackage[linesnumbered,ruled]{algorithm2e}

\newcommand{\mmat}[1]{\mathbf{#1}}
\newcommand{\mvec}[1]{\bm{#1}}
\newcommand{\mset}[1]{\mathcal{#1}}

\newcommand{\mtrans}{^\intercal}
\newcommand{\mbaseSet}[1]{\mathbb{#1}}
\newcommand{\msetBuilder}[2]{\{{#1}\ |\ {#2}\}}
\newcommand{\msetList}[1]{\{{#1}\}}

\newcommand{\mquant}[1]{|#1|}

\newcommand{\mfunc}[1]{\mathsf{#1}}

\newcommand{\mmod}{\ \textnormal{mod}\ }

\newcommand{\mComplexity}[1]{\mset{O}(#1)}

\newcommand{\plaDimension}{n}

\newcommand{\plaGenericVector}[2]{(#1_0, \dots, #1_{#2 - 1})}

\newcommand{\plaIteration}{\mvec{i}}
\newcommand{\plaIntraIteration}{\mvec{j}}
\newcommand{\plaInterIteration}{\mvec{k}}
\newcommand{\plaTiledIteration}{\mvec{i}^*}

\newcommand{\plaNode}{v}

\newcommand{\plaIntraIterationElement}{j}

\newcommand{\plaIterationSpace}{\mset{I}}
\newcommand{\plaIterationSpaceDefinition}{\plaIterationSpace \subseteq \mbaseSet{Z}^\plaDimension}
\newcommand{\plaIterationLongDefinition}{\plaIteration=\plaGenericVector{i}{\plaDimension}\mtrans \in \plaIterationSpace}

\newcommand{\plaEquation}[1][i]{S_{#1}}
\newcommand{\plaEquationSpace}{\mset{S}}
\newcommand{\plaEquationDomain}[1][i]{\mset{I}_{#1}}
\newcommand{\plaEquationSpaceDefinition}{\plaEquationSpace = \{\plaEquation[0], \plaEquation[1], \dots\}}
\newcommand{\plaOperation}[1][i]{\mfunc{f}_{#1}}

\newcommand{\plaWriteIndexingFunctionMatrix}[1][i]{\mmat{P}_{#1}}
\newcommand{\plaWriteIndexingFunctionOffset}[1][i]{\mvec{f}_{#1}}
\newcommand{\plaWriteIndexingFunction}[1][i]{\plaWriteIndexingFunctionMatrix[#1]\plaIteration + \plaWriteIndexingFunctionOffset[#1]}

\newcommand{\plaReadIndexingFunctionMatrix}[1][]{%
\if\relax\detokenize{#1}\relax
    \mmat{Q}%
\else
    \mmat{Q}_{#1}%
\fi
}
\newcommand{\plaReadIndexingFunctionOffset}[1][]{%
\if\relax\detokenize{#1}\relax
    \mvec{d}%
\else
    \mvec{d}_{#1}%
\fi
}
\newcommand{\plaReadIndexingFunction}[1][{i, j}]{\plaReadIndexingFunctionMatrix[#1]\plaIteration - \plaReadIndexingFunctionOffset[#1]}

\newcommand{\plaVariable}[1]{#1}

\newcommand{\plaVariableAt}[2]{\plaVariable{#1}[#2]}

\newcommand{\plaEquationDefinition}{\plaEquation[i]: \plaVariableAt{x_i}{\plaWriteIndexingFunction} = \plaOperation[i](\dots, \plaVariableAt{y_{i, j}}{\plaReadIndexingFunction}, \dots) \quad \textnormal{if } \plaIteration \in \plaEquationDomain[i]}

\newcommand{\plaIntra}{^{\plaIntraIteration}}
\newcommand{\plaInter}{^{\plaInterIteration}}

\newcommand{\plaIntraIterationSpace}{\mset{J}}

\newcommand{\plaInterIterationSpace}{\mset{K}}

\newcommand{\plaTiledIterationSpace}{\mset{I}^*}
\newcommand{\plaTiledIterationSpaceDefinition}{\plaTiledIterationSpace=\msetBuilder{\plaTiledIteration=(\plaIntraIteration\mtrans, \plaInterIteration\mtrans)\mtrans}{\plaIntraIteration \in \plaIntraIterationSpace \land \plaInterIteration \in \plaInterIterationSpace}}

\newcommand{\plaTiledIntraScheduleVector}{\mvec{\lambda}\plaIntra}
\newcommand{\plaTiledInterScheduleVector}{\mvec{\lambda}\plaInter}
\newcommand{\plaTiledInterScheduleVectorElement}{\lambda\plaInter}
\newcommand{\plaTiledScheduleVector}{\mvec{\lambda^*}}
\newcommand{\plaTiledScheduleVectorDefinition}{\mvec{\lambda^*} = (\plaTiledIntraScheduleVector, \plaTiledInterScheduleVector)}
\newcommand{\plaScheduleOperationStartTime}[1][i]{\sigma_{#1}}

\newcommand{\plaScheduleOperationExecutionTime}[1][i]{\delta_{#1}}
\newcommand{\plaScheduleOperationScheduleTime}{t}
\newcommand{\plaScheduleOperationScheduleTimeDefinition}{\plaScheduleOperationScheduleTime = \plaTiledScheduleVector\plaTiledIteration+\plaScheduleOperationStartTime}

\newcommand{\plaScheduleInitiationInterval}{\textnormal{II}\xspace}
\newcommand{\plaScheduleLocalLatency}{L_\textnormal{local}}
\newcommand{\plaScheduleLocalLatencyDefinition}{L_\textnormal{local} = \max_i\ (\plaScheduleOperationStartTime[i] + \plaScheduleOperationExecutionTime[i]) - \min_i\ \plaScheduleOperationStartTime[i] }

\newcommand{\plaConditionSpace}{\plaEquationDomain}
\newcommand{\plaConditionSpaceMatrix}{\mmat{A}_i}

\newcommand{\plaConditionSpaceOffset}{\mvec{b}_i}

\newcommand{\plaConditionSpaceLongDefinition}{\plaConditionSpace = \msetBuilder{\plaIteration \in \plaIterationSpace}{\plaConditionSpaceMatrix\plaIteration \ge \plaConditionSpaceOffset}}

\newcommand{\plaProgram}[1]{%
    \begin{tikzpicture}
        \foreach \eq/\lhs/\rhs/\cnd [count=\i] in {#1} {
            \node[anchor=east] at (0, -\baselineskip * \i) {${\plaEquation[\eq]}\colon$};
        }   
    \end{tikzpicture}%
    \begin{tikzpicture}
        \foreach \eq/\lhs/\rhs/\cnd [count=\i] in {#1} {
            \node[anchor=west] at (0, -\baselineskip * \i) {$\lhs$};
        }   
    \end{tikzpicture}%
    \begin{tikzpicture}
        \foreach \eq/\lhs/\rhs/\cnd [count=\i] in {#1} {
            \node[anchor=west] at (0, -\baselineskip * \i) {$ =\ \rhs$};
        }   
    \end{tikzpicture}%
    \begin{tikzpicture}
        \foreach \eq/\lhs/\rhs/\cnd [count=\i] in {#1} {
            \ifthenelse{\equal{\cnd}{-}}{
            }{%
                \node[anchor=west] at (0, -\baselineskip * \i) {$\textnormal{ if } \cnd$};
            }
        }   
    \end{tikzpicture}
}%

\newcommand{\plaCtrlConditionSpace}{\mset{C}}
\newcommand{\plaCtrlCondition}{c}
\newcommand{\plaCtrlConditionZeroDomain}{\plaIntraIterationSpace^0}
\newcommand{\plaCtrlConditionOneDomain}{\plaIntraIterationSpace^1}
\newcommand{\plaCtrlConditionDefinition}{\plaCtrlCondition_i = (\plaCtrlConditionZeroDomain_i, \plaCtrlConditionOneDomain_i)}

\newcommand{\plaCtrlPrime}{%
\plaCtrlConditionOneDomain_i \subseteq \plaCtrlConditionOneDomain_j \land
\plaCtrlConditionZeroDomain_i \subseteq \plaCtrlConditionZeroDomain_j}
\newcommand{\plaCtrlInversePrime}{%
\plaCtrlConditionZeroDomain_i \subseteq \plaCtrlConditionOneDomain_j \land
\plaCtrlConditionOneDomain_i \subseteq \plaCtrlConditionZeroDomain_j}

\newcommand{\plaCtrlPrimeReverse}{%
\plaCtrlConditionOneDomain_j \subseteq \plaCtrlConditionOneDomain_i \land
\plaCtrlConditionZeroDomain_j \subseteq \plaCtrlConditionZeroDomain_i}
\newcommand{\plaCtrlInversePrimeReverse}{%
\plaCtrlConditionZeroDomain_j \subseteq \plaCtrlConditionOneDomain_i \land
\plaCtrlConditionOneDomain_j \subseteq \plaCtrlConditionZeroDomain_i}

\newcommand{\plaCtrlConditionAdd}{%
(\plaCtrlConditionZeroDomain_i \cup \plaCtrlConditionZeroDomain_j, %
\plaCtrlConditionOneDomain_i \cup \plaCtrlConditionOneDomain_j)}
\newcommand{\plaCtrlConditionInverseAdd}{%
(\plaCtrlConditionOneDomain_i \cup \plaCtrlConditionZeroDomain_j, %
\plaCtrlConditionZeroDomain_i \cup \plaCtrlConditionOneDomain_j)}

\newcommand{\plaCtrlPrimeConditionSpace}{\mset{C}_\textnormal{prime}}

\newcommand{\plaCtrlCompability}{\plaCtrlConditionZeroDomain_i \cap \plaCtrlConditionOneDomain_j = \emptyset \land \plaCtrlConditionOneDomain_i \cap \plaCtrlConditionZeroDomain_j = \emptyset}
\newcommand{\plaCtrlInverseCompability}{\plaCtrlConditionZeroDomain_i \cap \plaCtrlConditionZeroDomain_j = \emptyset \land \plaCtrlConditionOneDomain_i \cap \plaCtrlConditionOneDomain_j = \emptyset}

\newcommand{\plaScheduleOperationWrap}{\plaScheduleOperationStartTime' = \plaScheduleOperationStartTime \mmod \plaScheduleInitiationInterval}
\newcommand{\plaScheduleOperationShiftDomain}{{\plaIntraIterationSpace_i}'}
\newcommand{\plaScheduleOperationShiftDomainDefinition}{
  \plaScheduleOperationShiftDomain =
  \msetBuilder{\plaIntraIteration'}{
    \exists \plaIntraIteration \in \plaIntraIterationSpace_i
    \ \ \plaTiledIntraScheduleVector\plaIntraIteration +
    \left\lfloor\frac{\plaScheduleOperationStartTime}{\plaScheduleInitiationInterval}\right\rfloor
    \cdot \plaScheduleInitiationInterval =
    \plaTiledIntraScheduleVector\plaIntraIteration'
  }
}

\newcommand{\plaCtrlConditionUnified}{{\mset{C}_\textnormal{unified}}}
\newcommand{\plaCtrlConditionPrimeAlgo}{%
\begin{algorithm}\small
\caption{Prime Control Condition}
\label{algo:ctrlPrime}
\KwIn{$\plaCtrlConditionSpace$}
\KwOut{$\plaCtrlPrimeConditionSpace$}
$\plaCtrlPrimeConditionSpace = \emptyset$\;
\ForEach{$\plaCtrlCondition_i = (\plaCtrlConditionZeroDomain_i, \plaCtrlConditionOneDomain_i) \in \plaCtrlConditionSpace$}{
    \ForEach{$\plaCtrlCondition_j = (\plaCtrlConditionZeroDomain_j, \plaCtrlConditionOneDomain_j) \in \plaCtrlPrimeConditionSpace$}{
        \If{$\plaCtrlPrime\lor\plaCtrlInversePrime$}{
            continue with next $\plaCtrlCondition_i$\;
        }
        \If{$\plaCtrlPrimeReverse\lor\plaCtrlInversePrimeReverse$}{
            $\plaCtrlPrimeConditionSpace = \plaCtrlPrimeConditionSpace \setminus \msetList{\plaCtrlCondition_j} \cup \msetList{\plaCtrlCondition_i}$\;
            continue with next $\plaCtrlCondition_i$\;
        }
    }
    $\plaCtrlPrimeConditionSpace = \plaCtrlPrimeConditionSpace \cup \msetList{\plaCtrlCondition_i}$\;
}
\Return{$\plaCtrlPrimeConditionSpace$}\;
\end{algorithm}
}
\newcommand{\plaCtrlConditionUnifiedAlgo}{%
\begin{algorithm}\small
\caption{Control Condition Unification}
\label{algo:ctrlUnified}
\KwIn{$\plaCtrlPrimeConditionSpace$}
\KwOut{$\plaCtrlConditionUnified$}
$\plaCtrlConditionUnified = \emptyset$\;
\ForEach{$\plaCtrlCondition_i = (\plaCtrlConditionZeroDomain_i, \plaCtrlConditionOneDomain_i) \in \textnormal{rand}(\plaCtrlPrimeConditionSpace)$}{
    \ForEach{$\plaCtrlCondition_j = (\plaCtrlConditionZeroDomain_j, \plaCtrlConditionOneDomain_j) \in \plaCtrlConditionUnified$}{
        \If{$\plaCtrlCompability$}{%
            $\plaCtrlConditionUnified = \plaCtrlConditionUnified \setminus \msetList{\plaCtrlCondition_j} \cup \{\plaCtrlConditionAdd\}$\;
            \vspace{-0.0cm}continue with next $\plaCtrlCondition_i$\;
        }
        \If{$\plaCtrlInverseCompability$}{
            $\plaCtrlConditionUnified = \plaCtrlConditionUnified \setminus \msetList{\plaCtrlCondition_j} \cup \{\plaCtrlConditionInverseAdd\}$\;
            \vspace{-0.0cm}continue with next $\plaCtrlCondition_i$\;
        }
    }
    $\plaCtrlConditionUnified = \plaCtrlConditionUnified \cup \msetList{\plaCtrlCondition_i}$\;
}
\Return{$\plaCtrlConditionUnified$}\;
\end{algorithm}
}

%% file: ctrl_paper.bib
@article{KoulMSTNZLSCMSDDCKFHNSTBDMTRBFHBHT23,
  author       = {Kalhan Koul and
                  Jackson Melchert and
                  Kavya Sreedhar and
                  Leonard Truong and
                  Gedeon Nyengele and
                  Keyi Zhang and
                  Qiaoyi Liu and
                  Jeff Setter and
                  Po{-}Han Chen and
                  Yuchen Mei and
                  Maxwell Strange and
                  Ross Daly and
                  Caleb Donovick and
                  Alex Carsello and
                  Taeyoung Kong and
                  Kathleen Feng and
                  Dillon Huff and
                  Ankita Nayak and
                  Rajsekhar Setaluri and
                  James Thomas and
                  Nikhil Bhagdikar and
                  David Durst and
                  Zachary Myers and
                  Nestan Tsiskaridze and
                  Stephen Richardson and
                  Rick Bahr and
                  Kayvon Fatahalian and
                  Pat Hanrahan and
                  Clark W. Barrett and
                  Mark Horowitz and
                  Christopher Torng and
                  Fredrik Kjolstad and
                  Priyanka Raina},
  title        = {{AHA:} An Agile Approach to the Design of Coarse-Grained Reconfigurable
                  Accelerators and Compilers},
  journal      = {{TECS}},
  number       = {2},
  year         = {2023},
  volume       = {22},
  pages        = {35:1--35:34},
  doi          = {10.1145/3534933},
}

@article{24_CGRA_Taxonomy,
    author = {Liu, Leibo and Zhu, Jianfeng and Li, Zhaoshi and Lu, Yanan and Deng, Yangdong and Han, Jie and Yin, Shouyi and Wei, Shaojun},
    title = {A Survey of Coarse-Grained Reconfigurable Architecture and Design: Taxonomy, Challenges, and Applications},
    journal = {Comput. Surv.},
    year = {2019},
    volume = {52},
    number = {6},
    pages = {118:1--118:39},
    doi = {10.1145/3357375},
}

@article{HannigReiche2014ACMTECS,
    author = {Hannig, Frank and Lari, Vahid and Boppu, Srinivas and Tanase, Alexandru and Reiche, Oliver},
    title = {{Invasive} {Tightly}-{Coupled} {Processor} {Arrays}: {A} {Domain}-{Specific} {Architecture}/{Compiler} {Co}-{Design} {Approach}},
    journal = {{TECS}},
    year = {2014},
    volume = {13},
    number = {4s},
    pages = {133:1--133:29},
    doi = {10.1145/2584660},
}

@article{nelis1988,
    author = {Nelis, Harry and Deprettere, Ed},
    title = {Automatic design and partitioning of systolic/wavefront arrays for {VLSI}},
    journal = {Circuits, Systems and Signal Processing},
    year = {1988},
    volume = {7},
    number = {2},
    pages = {pages 235--252}
}

@article{TanaseHannig2018ATECS,
    author = {Tanase, Alexandru and Witterauf, Michael and Teich, J{\"u}rgen and Hannig, Frank},
    title = {Symbolic {{Multi-Level Loop Mapping}} of {{Loop Programs}} for {{Massively Parallel Processor Arrays}}},
    journal = {{TECS}},
    year = {2018},
    volume = {17},
    number = {2},
    pages = {1--27},
    doi = {10.1145/3092952},
}

@article{WitteraufWHT21,
    author = {Michael Witterauf and Dominik Walter and Frank Hannig and J{\"{u}}rgen Teich},
    title = {Symbolic Loop Compilation for Tightly Coupled Processor Arrays},
    journal = {{TECS}},
    year = {2021},
    volume = {20},
    number = {5},
    pages = {49:1--49:31},
    doi = {10.1145/3466897},
}

@article{nickolls2008scalable,
  author = {Nickolls, John and Buck, Ian and Garland, Michael and Skadron, Kevin},
  title = {{Scalable Parallel Programming with CUDA}},
  journal = {Queue},
  year = {2008},
  volume = {6},
  number = {2},
  pages = {40--53},
  doi = {10.1145/1365490.1365500}
}

@article{lindholm2008nvidia,
  author       = {Erik Lindholm and
                  John Nickolls and
                  Stuart F. Oberman and
                  John Montrym},
  title        = {{NVIDIA} Tesla: {A} Unified Graphics and Computing Architecture},
  journal      = {Micro},
  year         = {2008},
  volume       = {28},
  number       = {2},
  pages        = {39--55},
  doi          = {10.1109/MM.2008.31}
}

@article{TirelliSAFDAMAP24,
  author       = {Cristian Tirelli and
                  Juan Sapriza and
                  Rub{\'{e}}n Rodr{\'{\i}}guez {\'{A}}lvarez and
                  Lorenzo Ferretti and
                  Beno{\^{\i}}t W. Denkinger and
                  Giovanni Ansaloni and
                  Jose Angel Miranda and
                  David Atienza and
                  Laura Pozzi},
  title        = {SAT-Based Exact Modulo Scheduling Mapping for Resource-Constrained CGRAs},
  journal      = {JETC},
  year         = {2024},
  volume       = {20},
  number       = {3},
  pages        = {8:1--8:26},
  doi          = {10.1145/3663675}
}

@incollection{TeichWitterauf2022Book,
    author = {Teich, J{\"u}rgen and Brand, Marcel and Hannig, Frank and Heidorn, Christian and Walter, Dominik and Witterauf, Michael},
    title = {{Invasive} {Tightly}-{Coupled} {Processor} {Arrays}},
    booktitle = {Invasive Computing},
    publisher = {FAU University Press},
    year = {2022},
    %isbn = {978-3-96147-571-1},
    doi = {10.25593/978-3-96147-571-1},
}

@inproceedings{26_CGRA_Overview,
    author = {Wijtvliet, Mark and Waeijen, Luc and Corporaal, Henk},
    title = {Coarse grained reconfigurable architectures in the past 25 years: Overview and classification},
    booktitle = {SAMOS},
    %location = {Agios Konstantinos, Samos Island, Greece},
    year = {2016},
    pages = {235-244},
    doi = {10.1109/SAMOS.2016.7818353},
}

@inproceedings{alpaca,
    author = {Walter, Dominik and Brand, Marcel and Heidorn, Christian and Witterauf, Michael and Hannig, Frank and Teich, Jürgen},
    title = {{ALPACA}: {An} {Accelerator} {Chip} for {Nested} {Loop} {Programs}},
    booktitle = {ISCAS},
    %location = {Singapur},
    year = {2024},
    pages = {1--5},
    doi = {10.1109/ISCAS58744.2024.10558549},
}

@inproceedings{BrandTeich2017IEEEMCSOC,
    author = {{Brand}, {Marcel} and {Hannig}, {Frank} and {Tanase}, {Alexandru} and {Teich}, {J\"urgen}},
    title = {Orthogonal Instruction Processing: {An} Alternative to Lightweight {VLIW} Processors},
    booktitle = {MCSoC},
    %location = {Seoul, South Korea},
    year = {2017},
    pages = {5--12},
    doi = {10.1109/MCSoC.2017.17},
}

@inproceedings{KisslerTeich2006IEEEFPT,
    author = {{Kissler},{Dmitrij} and {Hannig},{Frank} and {Kupriyanov},{Alexey} and {Teich},{J{\"u}rgen}},
    title = {{A} {Highly} {Parameterizable} {Parallel} {Processor} {Array} {Architecture}},
    booktitle = {FPT},
    %location = {Bangkok, Thailand},
    year = {2006},
    pages = {105--112},
    doi = {10.1109/FPT.2006.270293},
}

@inproceedings{WalterT21,
    author = {Dominik Walter and J{\"{u}}rgen Teich},
    title = {{LION}: {Real}-Time {I/O} Transfer Control for Massively Parallel Processor Arrays},
    booktitle = {MEMOCODE},
    %location = {Virtual Event},
    year = {2021},
    pages = {32--43},
    doi = {10.1145/3487212.3487349},
}

@inproceedings{WangKMMP19,
    author = {Bo Wang and Manupa Karunarathne and Aditi Kulkarni Mohite and Tulika Mitra and Li{-}Shiuan Peh},
    title = {{HyCUBE}: {A}{} 0.9{V} 26.4 {MOPS/mW},290 {pJ}/op, Power Efficient Accelerator for {IoT} Applications},
    booktitle = {A-SSCC},
    %location = {Macau, SAR, China},
    year = {2019},
    pages = {133--136},
    doi = {10.1109/A-SSCC47793.2019.9056954},
}

@inproceedings{WitteraufTeich2016ASAP,
    author = {Witterauf, Michael and Tanase, Alexandru and Hannig, Frank and Teich, J{\"u}rgen},
    title = {Modulo Scheduling of Symbolically Tiled Loops for Tightly Coupled Processor Arrays},
    booktitle = {ASAP},
    %location = {London, United Kingdom},
    year = {2016},
    pages = {58--66},
    doi = {10.1109/ASAP.2016.7760773},
}

@inproceedings{WangTOSAAPA25,
  author       = {Yuxuan Wang and
                  Cristian Tirelli and
                  Lara Orlandic and
                  Juan Sapriza and
                  Rub{\'{e}}n Rodr{\'{\i}}guez {\'{A}}lvarez and
                  Giovanni Ansaloni and
                  Laura Pozzi and
                  David Atienza},
  title        = {An MLIR-Based Compilation Framework for {CGRA} Application Deployment},
  booktitle    = {ARC},
  pages        = {33--50},
  year         = {2025},
  doi          = {10.1007/978-3-031-87995-1\_3},
}

@inproceedings{edgeCGRA,
author = {\'{A}lvarez, Rub\'{e}n Rodr\'{\i}guez and Denkinger, Beno\^{\i}t and Sapriza, Juan and Calero, Jos\'{e} Miranda and Ansaloni, Giovanni and Alonso, David Atienza},
title = {An Open-Hardware Coarse-Grained Reconfigurable Array for Edge Computing},
booktitle = {CF},
pages = {391–392},
year = {2023},
doi = {10.1145/3587135.3591437},
}

@inproceedings{das2018efficient,
  author       = {Satyajit Das and
                  Kevin J. M. Martin and
                  Philippe Coussy and
                  Davide Rossi and
                  Luca Benini},
  title        = {Efficient mapping of {CDFG} onto coarse-grained reconfigurable array
                  architectures},
  booktitle    = {ASP-DAC},
  pages        = {127--132},
  year         = {2017},
  doi          = {10.1109/ASPDAC.2017.7858308}
}

@inproceedings{BednaraHT02,
  author       = {Marcus Bednara and
                  Frank Hannig and
                  J{\"{u}}rgen Teich},
  title        = {Generation of Distributed Loop Control},
  booktitle    = {SAMOS},
  pages        = {154--170},
  year         = {2002},
  doi          = {10.1007/3-540-45874-3\_9},
}

@misc{polybench,
    %note = {Accessed: 20 Apr 2024},
    howpublished = {\url{https://web.cs.ucla.edu/~pouchet/software/polybench/}},
    year = {2012},
    author = {Louis-Noel Pouchet},
    title = {PolyBench/C: The Polyhedral Benchmark suite},
}

@thesis{Tei93,
    %isbn = {978-3-86111-701-8},
    type = {Dissertation},
    school = {Saarland University, Germany},
    title = {A Compiler for Application Specific Processor Arrays},
    pagetotal = {230 pp. Shaker Verlag},
    year = {1993},
    author = {J{\"{u}}rgen Teich},
}
